\documentclass[twocolumn,showpacs,preprintnumbers,amsmath,amssymb]{revtex4}

\usepackage{graphicx}
\usepackage{dcolumn}
\usepackage{bm}

\def\be{\begin{equation}}
\def\ee{\end{equation}}
\def\bea{\begin{eqnarray}}
\def\eea{\end{eqnarray}}

\begin{document} \preprint{\large\it }

\sloppy

\title{Robust manipulation of electron spin coherence in an ensemble of singly charged quantum dots}

\author{A. Greilich, M. Wiemann, F.~G.~G. Hernandez$^{\dag}$, D.~R. Yakovlev$^{\S}$,  I.~A. Yugova$^{\ddag}$, and M.
Bayer}

\affiliation{Experimentelle Physik II, Universit\"at Dortmund,
D-44221 Dortmund, Germany}

\author{A. Shabaev$^{\star}$ and Al.~L. Efros}

\affiliation{Naval Research Laboratory, Washington, DC 20375, USA}

\author{D. Reuter and A.~D. Wieck}

\affiliation{Angewandte Festk\"orperphysik, Ruhr-Universit\"at
Bochum, D-44780 Bochum, Germany}

\date{\today, robustcontrol-03-27-07-fin.tex}

\begin{abstract}
Using the recently reported mode locking effect \cite{Grei06} we
demonstrate a highly robust control of electron spin coherence in
an {\em ensemble} of (In,Ga)As quantum dots during the single spin
coherence time.  The spin precession in a transverse magnetic
field can be fully controlled up to 25\, K by the parameters of
the exciting pulsed laser protocol such as the pulse train
sequence, leading to adjustable quantum beat bursts in Faraday
rotation. Flipping of the electron spin precession phase was
demonstrated by inverting the polarization within a pulse doublet
sequence.
\end{abstract}

\pacs{72.25.Dc, 72.25.Rb, 78.47.+p, 78.55.Cr}

\maketitle

The spin of an electron in a quantum dot (QD) is an attractive
quantum bit candidate \cite{LossPRA98,ImamogluPRL99,Spintr,Yamamoto} due to
its favorable coherence properties
\cite{Grei06,Elz04,Kro04,Pet05}. As the interaction strength is rather small for direct spin
manipulation, the idea to swap spin into charge has been furbished
\cite{Elz04,Hanson05,Calarco03}. For example, the electron may be
converted into a charged exciton by optical injection of an
electron-hole pair \cite{Calarco03}, depending on the residual
electron's spin orientation, leading to distinctive polarization
selection rules.

\begin{figure}
  \centering
\includegraphics[width=7.5cm]{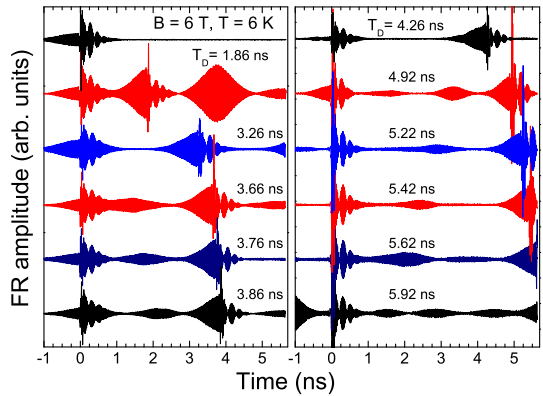}
\caption{Faraday rotation traces measured as function of delay
between probe and first pump pulse at time zero. A second pump
pulse was applied, delayed relative to the first one by $T_D$,
indicated at each trace. The top left trace gives the FR without
second pump.}
\end{figure}

The fundamental quantity regarding spin coherence is the
transverse relaxation time $T_2$. In a QD ensemble, this time is
masked by dephasing, mostly caused by dot-to-dot variations of the
spin dynamics. The dephasing time does not exceed 10 ns, much
shorter than $T_2$. This leads to the general believe that
manipulations ought to be performed on a single spin. Measurement
of a single electron spin polarization, however, also results in
dephasing due to temporal sampling of varying nuclear spin
configurations \cite{Sham,Liu}, as statistically significant
measurements on a single QD may require multiple repetition of the
experiment. The dephasing can be overcome by spin-echo techniques,
which give a single electron spin coherence time on the scale of
micro-seconds \cite{Pet05}. This long coherence time derived by
spin-echo is result of a refocusing of the electron spin and
possibly the nuclear spin configuration \cite{Sham}, and it is
viewed as an upper bound on the free-induction decay of spin
coherence \cite{Sham, CoishLoss}.

Recently, however, we have shown that mode locking of electron
spin coherence allows one to overcome the ensemble dephasing
\cite{Kurizki} and to measure the single electron spin relaxation
time $T_2$ without applying spin-echo refocusing \cite{Grei06}.
For monitoring the coherence, pump-probe Faraday rotation (FR)
measurements \cite{Kikkawa} on a QD ensemble were used: after
optical alignment of the spins normal to an external magnetic
field the electron spins precess about this field. Due to
precession frequency variations the ensemble phase coherence is
quickly lost. However, a periodic train of circularly polarized
pulses emitted by a mode-locked laser synchronizes those spin
precession modes, for which the precession frequency is a multiple
of the laser repetition rate. This synchronization leads to
constructive interference (CI) of these modes in the ensemble spin
polarization before arrival of each pump pulse (see Fig. 1, upper
left trace). The limit for spin mode locking is set by the single
electron spin coherence time which can last up to a few
microseconds \cite{Grei06} reaching the low bound on echo-like
decays \cite{Hanson}.

Here we develop a detailed understanding of the degree of control
which can be reached for the electron spin coherence in an
ensemble of singly charged QDs by exploiting the mode locking. For
this purpose trains of excitation pump pulse doublets were
designed to vary the phase synchronization condition (PSC) for
electron spin precession frequencies. The PSC selects a QD subset,
whose contribution to the ensemble spin polarization shows a well
controlled phase recovery. Variation of the pulse separation
results in tunable patterns of quantum oscillation bursts in
time-resolved FR, in good agreement with our calculation, which
rely on a newly developed theoretical model. This tailoring of
electron spin coherence is very robust, as the spin mode locking
is stable up to 25 K. For higher temperatures the coherence
amplitude decreases due to phonon-assisted scattering of holes
during the laser pulse excitation by which the spin coherence is
created.

The studied self-assembled (In,Ga)As/GaAs QDs were fabricated by
molecular beam epitaxy on a (001)-oriented GaAs substrate. The
sample contains 20 QD layers with a layer dot density of about
10$^{10}$\,cm$^{-2}$, separated by 60 nm wide barriers
\cite{Grei05}. For average occupation by a single electron per
dot, the structures were $n$-modulation doped 20 nm below each
layer with a Si-dopant density matching roughly the dot density.
The sample was held in the insert of an optical magneto-cryostat,
allowing temperature variation from $T = 6$ to 50~K.

FR with picosecond time resolution was used for studying the spin
dynamics: Thereby spin polarization along the growth direction
($z$-axis) is generated by a circularly polarized pump pulse
hitting the sample along $z$, and its precession in a transverse
magnetic field $B \leq 7$\,T along the $x$-axis is tested by the
rotation of the linear polarization of a probe pulse. For optical
excitation, a Ti-sapphire laser was used emitting pulses with a
duration of $\sim$1.5 ps (full width at half maximum of $\sim$1
meV) at 75.6 MHz repetition rate (corresponding to a repetition
period $T_R = 13.2$\,ns). The laser energy was tuned into
resonance with the QD ground state transition and the laser pulses
were split into pump and probe. The pump beam was split further
into two pulses with variable delay $T_D$ in between. The circular
polarization of the two pumps could be controlled individually.
For detecting the rotation angle of the probe beam linear
polarization, a homodyne technique was used.

Figure 1 shows FR traces excited by the two-pulse train with a
repetition period $T_R = 13.2$\,ns, in which both pulses have the
same intensity and polarization, and the delay between these
pulses $T_D$ was varied between $\sim T_R/7$ and $\sim T_R/2$. The
FR pattern varies strongly for the case when the delay time $T_D$
is commensurate with the repetition period $T_R$: $T_D= T_R/i$
with $i=2,3,...$, and for the case $T_D\neq T_R/i$. For
commensurability $T_D=T_R/i$, the FR signal shows strong periodic
bursts of quantum oscillations only at times equal to multiples of
$T_D$, as seen in the left panel of Fig. 1 for
$T_D=1.86$\,ns$\approx T_R/7$. Commensurability is also given to a
good approximation for delays $T_D=T_R/4\approx 3.26$\,ns and
$T_D=T_R/3\approx 4.26$\,ns.

For incommensurability of $T_D$ and $T_R$, $T_D\neq T_R/i$, the FR
signal shows bursts of quantum oscillations between the two pulses
of each pump doublet, in addition to the bursts outside of the
doublet. For example, one can see a single burst in the mid
between the pumps for $T_D=3.76$ and 5.22\,ns. Two bursts, each
equidistant from the closest pump and also equidistant from one
another, appear at $T_D=4.92~{\rm and}~5.62$\,ns. Three
equidistant bursts occur at $T_D=5.92$\,ns. Note also that the FR
amplitude before the second pump arrival is always significantly
larger than before the first pump for any $T_D$.

Although the time dependencies of the FR signal look very
different for commensurate and incommensurate $T_D$ and $T_R$, in
both cases they can be fully controlled by designing the
synchronization of electron spin precession modes in order to
reach CI of their contributions to the FR signal \cite{Grei06}. A
train of circularly polarized pump pulse singlets synchronizes
those spin precessions for which the precession frequency
satisfies the PSC \cite{Grei06,SEM}: $\omega_e=2\pi N /T_R$.
Then the electron spin undergoes an integer number, $N\gg 1$, of
full $2\pi$ rotations in the interval $T_R$ between the pump
pulses.

For a train of pump pulse doublets the PSC has to be extended to
account for the intervals $T_D$ and $T_R-T_D$ in the laser
excitation protocol \be \omega_e=2\pi NK/T_D=2\pi NL/(T_R -T_D)~,
\label{eq1} \ee where $K$ and $L$ are integers. On first glance
this condition imposes severe limitations on the $T_D$ values, for
which synchronization is obtained: \be T_D=[K/( K+L)] T_R ~,
\label{eq2} \ee which for $T_D<T_R/2$ leads to $K<L$. When Eq.
(\ref{eq2}) is satisfied, the contribution of synchronized
precession modes to the average electron spin polarization
$\overline{S}_z(t)$ is proportional to $-0.5\cos[N(2\pi K
t/T_D)]$. Summing over all relevant oscillations leads to CI of
their contributions with a period $T_D/K$ in time \cite{Grei06}.
The rest of QDs does not contribute to $\overline{S}_z(t)$ at
times longer than the ensemble dephasing time. The PSC Eq.
(\ref{eq1}) explains the position of all bursts in the FR signal
for commensurate and incommensurate ratios of $T_D$ and $T_R$. For
commensurability, $K\equiv 1$ and $T_D=T_R/(1+L)$ according to Eq.
(\ref{eq2}). In this case CIs should occur with period $T_D$ as
seen in Fig. 1 for $T_D=1.86$\,ns ($L= 6$).

For incommensurability of $T_D$ and $T_R$ the number of FR bursts
between the two pulses within a pump doublet and the delays at
which they appear can be tailored. There should be just one burst
between the pulses, when $K\equiv 2$, because then the CI must
have a period $T_D/2$. A single burst is seen in Fig. 1 for
$T_D=3.76$ and 5.22\,ns. The corresponding ratios $T_D/T_R$ are
0.285 and 0.395, respectively. At the same time Eq.(\ref{eq2})
gives a ratio $T_D/T_R=2/(L+2)$, which is equal to 0.285 and 0.4
for $L$=5 and 3, respectively, in good accord with experiment.

Next, two FR bursts are seen for $T_D=4.92$ and 5.62\,ns,
corresponding to $T_D/T_R \approx $0.372 and 0.426. The
corresponding CI period $T_D/3$ is reached for $K\equiv 3$. Then
from Eq.(\ref{eq2}) $T_D/T_R=3/(L+3)$, giving 0.375 and 0.429 for
$L=5$ and 4, respectively. Finally, the FR signal with
$T_D=5.92$\,ns ($T_D/T_R \approx $0.448) shows three FR bursts
between the two pumps. The CI period $T_D/4$ is obtained for
$K\equiv 4$, for which Eq.(\ref{eq2}) gives
$T_D/T_R=4/(L+4)\approx 0.444$ with $L=5$. Obviously good general
agreement between experiment and theory is established,
highlighting the high flexibility of the laser protocol. In turn,
this understanding can be used to induce FR bursts at wanted
delays $T_D /K$, so that at these times further coherent
manipulation of all electron spins involved in the burst is
facilitated.

However, the question arises how accurate condition Eq.
(\ref{eq2}) for the $T_D/T_R$ ratio must be fulfilled to reach
phase synchronization. Formally, one can find for any arbitrary
$T_D/T_R$ large $K$ and $L$ values such that Eq. (\ref{eq2}) is
satisfied with high accuracy. But the above analysis shows, that
only the smallest of all available $L$ leads to PSC matching.
Experimentally, the facilities to address this point are limited,
as the largest $T_D$ for which FR signal can be measured are
delays around 5 ns between the two pumps. For larger delays the FR
bursts shift out of the scanning range. For short $T_D$, on the
other hand, the bursts are overlapping with the FR signal from the
pump pulses.

\begin{figure}
\includegraphics[width=6.5cm]{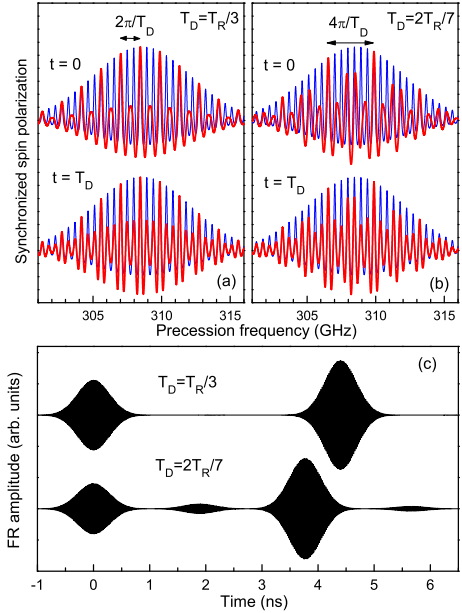}
\caption{(a,b): Spectra of electron spin precession modes,
$-\overline{S}_z(t)$, which are phase synchronized by the
two-pulse train calculated for $T_D=T_R/3$ and $T_D=2T_R/7$ at the
moments of first ($t=0$) and second ($t=T_D$) pulse arrival (red).
Single-pump spectra are shown in blue. (c): FR traces calculated
for two ratios of $T_D/T_R$. Laser pulse area $\Theta=\pi$.
$T_R=13.2$\,ns. Electron $g$-factor $\mid g_e \mid = 0.57$ and its
dispersion $\Delta g=0.005$. $B=6$~T.}
\end{figure}

To answer this question, we have modeled the FR signal for
commensurate and incommensurate ratios of $T_D$ and $T_R$. Figure
2 shows the results together with spectra of synchronized spin
precession modes (SSPM) at the moment of the first and second pump
pulse arrival. The SSPM were calculated similar those induced by a
single pulse train \cite{Grei06}. Figure 2(a) gives the SSPM for
commensurate $T_D=T_R/3$ superimposed on the SSPM created by a
single pulse train with the same $T_R$. Panel (c) shows the FR
signal created by such a two pulse train. The SSPM for the
considered strong excitation are considerably broadened and
contain modes for which $\omega_e=2\pi M/T_D=2\pi 3M/T_R$ with
integer $M$, which coincide with each third mode created by a
single pulse train. However, the SSPM given by $\omega_e=2\pi
N/T_R$, which do not satisfy the PSC for a two pulse train, are
not completely suppressed, because the train synchronizes the
electron spin precession in some frequency range around the PSC.
One sees also, that at $t=0$ the two pulse train leads to a
significant alignment of electron spins opposite to the direction
of spins satisfying the PSC. This "negative" alignment decreases
the CI magnitude and therefore the FR signal before the first
pulse arrival, and is also responsible for a significantly larger
magnitude of the FR signal before the second pulse arrival [see
Figs. 1  and 2 (c)].

For incommensurate ratios of $T_D$ and $T_R$ the SSPM  become much
more complex. Still we are able to recover the modes which satisfy
the PSC at the pulse arrival times. In Fig. 2 (b) we show the SSPM
at $t=0$ and $t=T_D$ for $T_D=2T_R/7$ ($K=2$, $L=5$), where the
arrows indicate the frequencies which satisfy the PSC for the two
pulse train. Only a small number of such modes fall within the
average distribution of electron spin precession modes, because
the distance between the PSC modes is proportional to $2\pi
K/T_D=2\pi(K+L)/T_R$. The diluted spectra of PSC modes for
incommensurability decrease the magnitude of the FR bursts between
the pump pulses, in accord with experiment. This shows, that
although any ratio of $T_D/T_R$ can be satisfied by large $K$ and
$L$, the FR signal between the pulses should be negligibly small
in this case. Consequently, not any ratio of $T_D/T_R$ leads to
pronounced FR bursts.

\begin{figure}
\includegraphics[width=6.5cm]{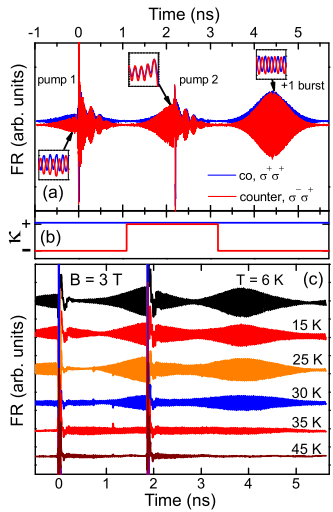}
\caption{(a): Faraday rotation traces in the co-circularly (blue
traces) or counter-circularly (red traces) polarized two pump
pulse experiments measured for $T_D=2.2$\,ns and $B$ = 6\,T.
Insets give close-ups showing the relative sign, $\kappa$, of the
FR amplitude between the two traces. $\kappa$ is plotted in (b) vs
time. (c): Effect of temperature on the FR amplitude in
two-pump-pulse experiment. $T_D=1.88$\,ns.}
\end{figure}

To obtain further insight into the tailoring of electron spin
coherence, which can be reached by a two-pulse train, we have
turned from co- to counter-circularly polarized pumps. The delay
between pumps $T_D$ was fixed at $T_R/6\approx 2.2$\,ns. The time
dependencies of the corresponding FR signals are similar, as shown
in Fig. 3. Besides the two FR bursts directly connected to the
pump pulses, one sees a burst $+1$ due to CI of spin synchronized
modes. The insets in Fig. 3 (a) show closeups of the different FR
bursts. The sign, $\kappa$, of the FR amplitude for the
counter-circular configuration undergoes $2T_D$-periodic changes
in time relative to the co-circular case, as seen in Fig. 3 (b),
which demonstrates optical switching of the electron spin
precession phase by $\pi$ in an {\em ensemble} of QDs.

The observed effect of sign reversal is well described by our
model. Let us consider first a two-pulse train with delay time
$T_D= T_R/2$ for which the two pumps are counter-circularly
polarized. In this case an electron spin can be synchronized only
if at the moment of pulse arrival it has an orientation opposite
to the orientation at the previous pulse. This leads to the PSC
$\omega_e=2\pi(N+1/2)/T_D$. The contribution of such precession
modes to the electron spin polarization is proportional to
$\cos[2\pi (N+1/2)t/T_D)]=\cos(2\pi Nt/T_D)\cos(\pi
t/T_D)-\sin(2\pi Nt/T_D)\sin(\pi t/T_D)$. Summing these
contributions, only the first term gives a CI, whose modulus has
period $T_D$, while the sign of $\cos(\pi t/T_D)$ changes with
period $2T_D$.  Only each third of the precession frequencies can
be synchronized by a counter-circularly polarized two pulse train
when the delay time is $T_R/6$ as in our experiment. However, the
corresponding PSC has the same dependence on $T_D$. The CI modulus
also has period $T_D$ and its sign changes with period $2T_D$. The
relative sign of the FR amplitude for the counter- and
co-circularly case, $\kappa={\rm sgn}\{\cos[\pi t/T_D]\}$, is in
accord with the experimental dependence in Fig. 3 (b).

The CIs of the electron spin contributions can be seen only as
long as the coherence of the electron spins is maintained. In this
respect the temperature stability of the CI is especially
important. Fig. 3 (c) shows FR traces in a two-pump-pulse
configuration with $T_D$ = 1.88 ns at different temperatures. For
both positive and negative delays, the FR amplitude at a fixed
delay is about constant for temperatures up to 25 K, irrespective
of slight variations which might arise from changes in the phase
synchronization of QD subsets. Above 30 K a sharp drop occurs,
which can be explained by thermally activated destruction of the
spin coherence.

The electron spin coherence in charged QDs is initiated by
generation of a superposition of an electron and a charged exciton
state by resonant pump pulses \cite{Grei05,Kenn06}. The
simultaneous decrease of the FR magnitude before each pump pulse
and afterwards (when the CI signal is controlled by the excitation
pulse) suggests that the coherence at elevated temperatures is
lost already during its generation. The 30\,K temperature
threshold corresponds to an activation energy of $\sim$2.5 meV.
This energy may be assigned only to the splitting between the two
lowest confined hole levels, because the electron level splitting
dominates the 20 meV splitting between p- and s-shell emission in
photoluminescence and is much larger than 2.5 meV. The decoherence
of the hole spin results from two phonon scattering, which is
thermally activated and should occur on a sub-picosecond time
scale, i.e. within the laser pulse \cite{Takaga}. The fast
decoherence of the hole spin at $T>30$~K suppresses formation of
the electron-trion superposition state. ps-pulses as used here are
therefore not sufficiently short for initialization of the
superposition and creation of a long-lived electron spin
coherence.

In summary, we have demonstrated that the mode-locking effect
allows a far-reaching control of electron spin coherence in QD
ensembles during the spin coherence time of microseconds
\cite{Grei06}. Two-pulse train mode-locking selects QD subsets
which give a non-dephasing contribution to the ensemble spin
precession. The technique shows remarkable stability with respect
to temperature increase up to 25 K, a property which is important
for utilizing it in quantum information processing. The robustness
of this control technique is provided by the dispersion of the
spin precession frequencies in the QD ensemble.

{\bf Acknowledgments.} This work was supported by the BMBF program
nanoquit, the DARPA program QuIST, the ONR, the DFG (FOR485) and
FAPESP.

\end{document}